# On an Algorithm for Isomorphism-Free Generations of Combinatorial Objects

**Krasimir Yordzhev**

Faculty of Mathematics and Natural Sciences
South-West University, Blagoevgrad, Bulgaria
*yordzhev@swu.bg*

**Abstract**: *In the work are defined the concepts semi-canonical and canonical binary matrix. What is described is an algorithm solving the combinatorial problem for finding the semi-canonical matrices in the set $\Lambda_n^k$ consisting of all $n \times n$ binary matrices having exactly $k$ 1's in every row and every column without perambulating all elements. In the described algorithm bitwise operations are substantially used. In this way it becomes easier to find the solution to the problem for receiving one representative from every equivalence class regarding the introduced in the article equivalence relation in the set $\Lambda_n^k$. The last problem is equivalent to the problem for finding all canonical matrices in $\Lambda_n^k$.*

## 1. INTRODUCTION

A big class of programming problems has the following general formulation:

**Problem 1.** *The set $\mathcal{M}$ is given. Compose a computer program receiving the set $\delta(\mathcal{M})$ from all elements of $\mathcal{M}$ possessing given properties. It is desirable to do this without perambulating all elements of the set $\mathcal{M}$.*

Variation of the problem above can be formulated in the following way:

**Problem 2.** *The set $\mathcal{M}$ is given and an equivalence relation is defined $\sim$ in $\mathcal{M}$. Compose a computer program receiving exactly one representative of every equivalence class regarding $\sim$, i.e. construct the factor-set $\mathcal{M}_{/\sim}$. Moreover without the necessity to perambulate all elements of $\mathcal{M}$.*

It is very often necessary to solve problems which solve simultaneously problems 1 and 2, i.e. to find the intersection $\delta(\mathcal{M}) \cap \mathcal{M}_{/\sim}$ of two sets, the first of which solves problem 1, and the second – problem 2.

We presume that there is a procedure which with a given element of $\mathcal{M}$ answers in the affirmative or in the negative whether this element belongs or respectively does not belong to $\delta(\mathcal{M}) \cap \mathcal{M}_{/\sim}$. The solution to such a problem in real time is complicated if the set $\mathcal{M}$ is too big, i.e. if its cardinality is an exponential function of one or more parameters. In this case the time necessary to perambulate all elements of the set and to check for every element whether it possesses the necessary properties is inefficiently large during the growth of the parameters.

One solution to the problems described above, which significantly increases the efficiency of the computer program, is to find a set $\mathcal{K} \subset \mathcal{M}$, such that $|\mathcal{K}| < |\mathcal{M}|$ and $\delta(\mathcal{M}) \cap \mathcal{M}_{/\sim} \subseteq \mathcal{K} \subset \mathcal{M}$, and the problem for finding $\mathcal{K}$ is algorithmically an easily solved problem. Then we will check whether they possess the given properties not for all elements of $\mathcal{M}$, but only for the elements of $\mathcal{K}$. The ideal case is when $\mathcal{K} = \delta(\mathcal{M}) \cap \mathcal{M}_{/\sim}$. Unfortunately this is not always an easy problem from an algorithmic point of view. One criterion for efficiency may be the fraction $\frac{|\mathcal{K}|}{|\mathcal{M}|}$, where the efficiency is inversely proportional to the value of this fraction. In the current work we will use this approach.

The algorithms solving the formulated problem are known as *algorithms for isomorphism-free generations of combinatorial objects*. Different approaches to creating such algorithms are discussed in [2].

The aim of the present article is to use the bitwise operations [6]-[8] in order to solve a problem from this class. The work is a continuation and addition to [7].

## 2. SEMI-CANONICAL AND CANONICAL BINARY MATRICES

*Binary* (or *boolean*, or *(0,1)-matrix*) is called a matrix whose elements belong to the set $\mathcal{B} = \{0,1\}$.

Let $n$ and $m$ be positive integers. With $\mathcal{B}_{n \times m}$ we will denote the set of all $n \times m$ binary matrices, while with $\mathcal{B}_n = \mathcal{B}_{n \times n}$ we will denote the set of all square $n \times n$ binary matrices.

**Definition 1.** 3 Let $A \in \mathcal{B}_{n \times m}$. With $r(A)$ we will denote the ordered $n$-tuple

$$\langle x_1, x_2, \ldots, x_n \rangle,$$

where $0 \leq x_i \leq 2^m - 1$, $i = 1, 2, \ldots n$ and $x_i$ is the integer written in binary notation with the help of the $i$-th row of $A$.

Similarly with $c(A)$ we will denote the ordered $m$-tuple

$$\langle y_1, y_2, \ldots, y_m \rangle,$$

where $0 \leq y_j \leq 2^n - 1$, $j = 1, 2, \ldots m$ and $y_j$ is the integer written in binary notation with the help of the $j$-th column of $A$.

We consider the sets:
$$\begin{aligned}\mathcal{R}_{n \times m} &= \{\langle x_1, x_2, \ldots, x_n \rangle \mid 0 \leq x_i \leq 2^m - 1, i = 1, 2, \ldots n\} \\ &= \{r(A) \mid A \in \mathcal{B}_{n \times m}\}\end{aligned}$$

and

$$\begin{aligned}\mathcal{C}_{n \times m} &= \{\langle y_1, y_2, \ldots, y_m \rangle \mid 0 \leq y_j \leq 2^n - 1, j = 1, 2, \ldots m\} \\ &= \{c(A) \mid A \in \mathcal{B}_{n \times m}\}\end{aligned}$$

With "<" we will denote the lexicographic orders in $\mathcal{R}_{n \times m}$ and in $\mathcal{C}_{n \times m}$.

It is easy to see that in definition 1 two mappings are described:

$$r : \mathcal{B}_{n \times m} \to \mathcal{R}_{n \times m}$$

and

$$c : \mathcal{B}_{n \times m} \to \mathcal{C}_{n \times m},$$

which are bijective and therefore

$$\mathcal{R}_{n \times m} \cong \mathcal{B}_{n \times m} \cong \mathcal{C}_{n \times m}.$$

In [3] it is proven that the representation of the elements of $\mathcal{B}_n$ using ordered $n$-tuples of natural numbers leads to the creation of faster and memory-saving algorithms.

**Definition 2.** 4 Let $A \in \mathcal{B}_{n \times m}$,

$$r(A) = \langle x_1, x_2, \ldots, x_n \rangle,$$
$$c(A) = \langle y_1, y_2, \ldots, y_m \rangle.$$

We will call the matrix $A$ *semi-canonical*, if

$$x_1 \leq x_2 \leq \cdots \leq x_n$$

and

$$y_1 \leq y_2 \leq \cdots \leq y_m.$$

**Proposition 1.** 5 *Let $A = [a_{ij}] \in \mathcal{B}_{n \times m}$ be a semi-canonical matrix. Then there exist integers $i$, $j$, such that $1 \leq i \leq n$, $1 \leq j \leq m$ and*

$$a_{11} = a_{12} = \cdots = a_{1j} = 0, \quad a_{1j+1} = a_{1j+2} = \cdots = a_{1m} = 1, \quad (1)$$

$$a_{11} = a_{21} = \cdots = a_{i1} = 0, \quad a_{i+11} = a_{i+21} = \cdots = a_{n1} = 1. \quad (2)$$

Proof. Let

$$r(A) = \langle x_1, x_2, \ldots x_n \rangle$$

and

$$c(A) = \langle y_1, y_2, \ldots y_m \rangle.$$

We assume that there exist integers $p$ and $q$, such that $1 \leq p < q \leq m$, $a_{1p} = 1$ and $a_{1q} = 0$. But in this case $y_p > y_q$, which contradicts the condition for semi-canonicity of the matrix $A$. We have proven (1). Similarly we prove (2) as well. □

Let $n$ be a positive integer. With

$$\mathcal{P}_n \subset \mathcal{B}_n = \mathcal{B}_{n \times n}$$

we will denote the set of all *permutation matrices*, i.e. the set of all $n \times n$ binary matrices having exactly one 1 in every row and every column. The isomorphism

$$\mathcal{P}_n \cong \mathcal{S}_n,$$

is true, where with $\mathcal{S}_n$ we have denoted the symmetric group, i.e. the group of all one-to-one mappings of the set

$$[n] = \{1, 2, \ldots n\}$$

in itself.

As it is well known [4], [5], the multiplication of an arbitrary real or complex matrix $A$ from the left with a permutation matrix (if the multiplication is possible) leads to dislocation of the rows of the matrix $A$. The multiplication of $A$ from the right with a permutation matrix leads to the dislocation of the columns of $A$.

**Definition 3.** 6 Let $A, B \in \mathcal{B}_{n \times m}$. We will say that the matrices $A$ and $B$ are equivalent and we will write

$$A \sim B, \qquad (3)$$

if there exist permutation matrices $X \in \mathcal{P}_n$ and $Y \in \mathcal{P}_m$, such that

$$A = XBY \qquad (4)$$

In other words $A \sim B$ if $A$ is received from $B$ after dislocation of some of the rows and the columns of $B$.

Obviously the introduced relation is an equivalence relation.

With
$$\mathcal{T}_n \subset \mathcal{P}_n$$
we denote the set of all *transpositions* in $\mathcal{P}_n$, i.e. the set of all $n \times n$ permutation matrices, which multiplying from the left an arbitrary $n \times m$ matrix swaps the places of exactly two rows, while multiplying from the right an arbitrary $k \times n$ matrix swaps the places of exactly two columns.

**Theorem 1. 7** *Let $A$ be an arbitrary matrix from $\mathcal{B}_{n \times m}$. Then:*

a) *If*
$$X_1, X_2, \cdots, X_s \in \mathcal{T}_n$$
*are such that*
$$r(X_1 X_2 \ldots X_s A) < r(X_2 X_3 \ldots X_s A) < \cdots < r(X_s A) < r(A),$$
*then*
$$c(X_1 X_2 \ldots X_s A) < c(A).$$

b) *If*
$$Y_1, Y_2, \cdots, Y_t \in \mathcal{T}_m$$
*are such that*
$$c(Y_1 Y_2 \ldots Y_t A) < c(Y_2 Y_3 \ldots Y_t A) < \cdots < c(X_t A) < r(A),$$
*then*
$$r(Y_1 Y_2 \ldots Y_t A) < r(A).$$

Proof. a) Induction by $s$. Let $s = 1$ and let $X \in \mathcal{T}_n$ be a transposition which multiplying an arbitrary matrix $A = [a_{ij}] \in \mathcal{B}_{n \times m}$ from the left swaps the places of the rows of $A$ with numbers $u$ and $v$ ($1 \leq u < v \leq n$), while the remaining rows stay in their places. In other words if

$$A = \begin{bmatrix} a_{11} & a_{12} & \cdots & a_{1r} & \cdots & a_{1m} \\ a_{21} & a_{22} & \cdots & a_{2r} & \cdots & a_{2m} \\ \vdots & \vdots & & \vdots & & \vdots \\ a_{u1} & a_{u2} & \cdots & a_{ur} & \cdots & a_{um} \\ \vdots & \vdots & & \vdots & & \vdots \\ a_{v1} & a_{v2} & \cdots & a_{vr} & \cdots & a_{vm} \\ \vdots & \vdots & & \vdots & & \vdots \\ a_{n1} & a_{n2} & \cdots & a_{nr} & \cdots & a_{nm} \end{bmatrix}$$

then

$$XA = \begin{bmatrix} a_{11} & a_{12} & \cdots & a_{1r} & \cdots & a_{1m} \\ a_{21} & a_{22} & \cdots & a_{2r} & \cdots & a_{2m} \\ \vdots & \vdots & & \vdots & & \vdots \\ a_{v1} & a_{v2} & \cdots & a_{vr} & \cdots & a_{vm} \\ \vdots & \vdots & & \vdots & & \vdots \\ a_{u1} & a_{u2} & \cdots & a_{ur} & \cdots & a_{um} \\ \vdots & \vdots & & \vdots & & \vdots \\ a_{n1} & a_{n2} & \cdots & a_{nr} & \cdots & a_{nm} \end{bmatrix},$$

where $a_{ij} \in \{0,1\}$, $1 \leq i \leq n$, $1 \leq j \leq m$.

Let
$$r(A) = \langle x_1, x_2, \ldots, x_u, \ldots, x_v, \ldots, x_n \rangle.$$
Then
$$r(XA) = \langle x_1, x_2, \ldots, x_v, \ldots, x_u, \ldots, x_n \rangle.$$

Since $r(XA) < r(A)$ then according to the properties of the lexicographic order $x_v < x_u$. According to definition 1 the representation of $x_u$ and $x_v$ in binary notation with an eventual addition if necessary with unessential zeroes in the beginning is respectively as follows:

$$x_u = a_{u1} a_{u2} \cdots a_{um},$$

$$x_v = a_{v1} a_{v2} \cdots a_{vm}.$$

Since $x_v < x_u$, then there exists an integer $r \in \{1, 2, \ldots, m\}$, such that $a_{uj} = a_{vj}$ when $j < r$, $a_{ur} = 1$ and $a_{vr} = 0$.

Let
$$c(A) = \langle y_1, y_2, \ldots, y_m \rangle$$
and
$$c(XA) = \langle z_1, z_2, \ldots, z_m \rangle.$$

Then $y_j = z_j$ when $j < r$, while the representation of $y_r$ and $z_r$ in binary notation with an eventual addition if necessary with unessential zeroes in the beginning is respectively as follows:

$$y_r = a_{1r} a_{2r} \cdots a_{u-1r} a_{ur} \cdots a_{vr} \cdots a_{nr},$$

$$z_r = a_{1r} a_{2r} \cdots a_{u-1r} a_{vr} \cdots a_{ur} \cdots a_{nr}.$$

Since $a_{ur} = 1$, $a_{vr} = 0$, then $z_r < y_r$, whence it follows that $c(XA) < c(A)$.

We assume that for every $s$-tuple of transpositions $X_1, X_2, \ldots, X_s \in \mathcal{T}_n$ and for every matrix $A \in \mathcal{B}_{n \times m}$ from

$$r(X_1 X_2 \ldots X_s A) < r(X_2 \cdots X_s A) < \cdots < r(X_s A) < r(A)$$

it follows that

$$c(X_1 X_2 \ldots X_s A) < c(A)$$

and let $X_{s+1} \in \mathcal{T}_n$ be such that

$$r(X_1 X_2 \ldots X_s X_{s+1} A) < r(X_2 \cdots X_{s+1} A) < \cdots < r(X_{s+1} A) < r(A).$$

According to the induction assumption

$$c(X_{s+1} A) < c(A).$$

We put

$$A_1 = X_{s+1} A.$$

According to the induction assumption from

$$r(X_1 X_2 \ldots X_s A_1) < r(X_2 \cdots X_s A_1) < \cdots < r(X_s A_1) < r(A_1)$$

it follows that

$$c(X_1 X_2 \cdots X_s X_{s+1} A) = c(X_1 X_2 \cdots X_s A_1) < c(A_1) =$$
$$= c(X_{s+1} A) < c(A),$$

with which we have proven a).
b) is proven similarly to a). □

**Corollary 1. 8***Let $A \in \mathcal{B}_{n \times m}$. Then $r(A)$ is a minimal element about the lexicographic order in the set $\{r(B) \mid B \sim A\}$ if and only if $c(A)$ is a minimal element about the lexicographic order in the set $\{c(B) \mid B \sim A\}$.*

Corollary 1 gives us grounds to formulate the following definition:

**Definition 4.** We will call the matrix $A \in \mathcal{B}_{n \times m}$ *canonical matrix*, if $r(A)$ is a minimal element about the lexicographic order in the set $\{r(B) \mid B \sim A\}$, and therefore $c(A)$ is a minimal element about the lexicographic order in the set $\{c(B) \mid B \sim A\}$.

From definition 4 follows that in every equivalence class about the relation "$\sim$" (definition 3) there exists only one canonical matrix.

If a matrix $A \in \mathcal{B}_{n \times m}$ is a canonical matrix, then it is easy to see that $A$ is a semi-canonical matrix, but as we will see in the next example the opposite statement is not always true.

**Example 1. 9**We consider the matrices:

$$A = \begin{bmatrix} 0 & 1 & 1 & 1 \\ 1 & 0 & 0 & 1 \\ 1 & 0 & 1 & 0 \\ 1 & 1 & 1 & 0 \end{bmatrix} \text{ and } B = \begin{bmatrix} 0 & 0 & 1 & 1 \\ 0 & 1 & 0 & 1 \\ 1 & 1 & 0 & 1 \\ 1 & 1 & 1 & 0 \end{bmatrix}$$

It is not difficult to see that $A \sim B$. Furthermore

$$r(A) = \langle 7, 9, 10, 14 \rangle,$$
$$c(A) = \langle 7, 9, 11, 12 \rangle,$$
$$r(B) = \langle 3, 5, 13, 14 \rangle,$$
$$c(B) = \langle 3, 7, 9, 19 \rangle.$$

Therefore $A$ and $B$ are two equivalent to each other semi-canonical matrices. □

From example 1 follows that in an equivalence class it is possible to exist more than one semi-canonical element.

Let $\mathcal{W}$ be an arbitrary subset of $\mathcal{B}_n$, such that if $A \in \mathcal{W}$ and $B \sim A$, then $B \in \mathcal{W}$. Then obviously there exists only one canonical matrix in every equivalence class in the factor-set $\mathcal{W}_{/\sim}$. Therefore the number of the canonical matrices $\mathcal{W}$ will give us the cardinality of the factor-set $\mathcal{W}_{/\sim}$.

## 3. FORMULATION OF THE PROBLEM

**Definition 5. 10**The square $n \times n$ binary matrices in every row and every column on which there exist exactly $k$ 1's we will call $\Lambda_n^k$-matrices, where with $\Lambda_n^k$ we will also denote the set of these matrices.

Let $n$ and $k$ be positive integers. We consider the sets:

$\mathcal{B}_n$ - the set of all square $n \times n$ binary matrices;

$\Lambda_n^k$ - the set of all square $n \times n$ binary matrices having exactly $k$ 1's in every row and every column;

$\mathcal{U}_n = \{A \in \mathcal{B}_n \mid r(A) = \langle x_1, x_2, \ldots, x_n \rangle, x_1 \le x_2 \le \cdots \le x_n\} \subset \mathcal{B}_n$;

$\mathcal{V}_n = \{A \in \mathcal{U}_n \mid c(A) = \langle y_1, y_2, \ldots, y_n \rangle, y_1 \le y_2 \le \cdots \le y_n\} \subset \mathcal{U}_n$;

$\mathcal{Z}_n^k = \Lambda_n^k \cap \mathcal{V}_n$ - the set of all semi-canonical matrices in $\Lambda_n^k$.

Let us consider the following problem:

**Problem 3.11** *Describe and implement an algorithm receiving the set $\mathcal{Z}_n^k$.*

By solving problem 3 we will facilitate to a minimum the solution to the problem for finding $\mu(n,k) = \left| \Lambda^k_{n/\sim} \right|$, i.e. the number of the equivalence classes according to definition 3 relation. The problem for finding the number $\mu(n,k)$ of the equivalence classes for every $n$ and $k$ is an open scientific problem. We will solve problem 3 by substantially using the properties of the bitwise operations with the aim of increasing the efficiency of the algorithm created by us.

## 4. Description and implementation of the algorithm by using bitwise operations

In the description of algorithms we will use the C++ programming language.

As it is well known, there are exactly $2^n$ nonnegative integers, which are presented with no more than $n$ digits in binary notation. We need to select all of them, which have exactly $k$ 1's in binary notation. Their number is $\binom{n}{k} \ll 2^n$. We will describe an algorithm that directly receives the necessary elements without checking whether any integer $m \in [0, 2^n - 1]$ satisfies the conditions. We will remember the result in the array p[] of size $c = \binom{n}{k}$. Moreover, the obtained array will be sorted in ascending order and there are no duplicate elements. The algorithm is based on the fact that the set of all ordered $m$-tuples

$$\mathcal{B}^m = \{\langle b_1, b_2, \ldots, b_m \rangle \mid b_i \in \mathcal{B} = \{0,1\}\},$$

$i = 1, 2, \ldots, m$, $m = 1, 2, \ldots, n$, is partitioned into two disjoint subsets

$$\mathcal{B}^m = M_1 \cup M_2, \quad M_1 \cap M_2 = \varnothing,$$

where

$$M_1 = \{\langle b_1, b_2, \ldots, b_m \rangle \mid b_1 = 0\}$$

and

$$M_2 = \{\langle b_1, b_2, \ldots, b_m \rangle \mid b_1 = 1\}.$$

The described below recursive algorithm will use the bitwise operations.

```
void DataNumb(int p[], unsigned int n, int k, int& c)
{
    if (k<=0)
    {
        c = 1;
        p[0] = 0;
    }
    else if (k==n)
    {
        c = 1;
        p[0] = (1<<n)-1;    // p[0] = 2^n - 1
    }
    else
    {
        int p1[10000], p2[10000];
        int c1, c2;
        DataNumb(p1, n-1, k, c1);
        DataNumb(p2, n-1, k-1, c2);
        c = c1+c2;
        for (int i=0; i<c1; i++)
            p[i] = p1[i];
        for (int i=0; i<c2; i++)
            p[c1+i] = p2[i] | 1<<(n-1);
    }
}
```

We will also use bitwise operations in constructing the next two functions.

The function int n_tuple(int[], int, int, int) gets all $t = \binom{n+k-1}{k}$ (combinations with repetitions) ordered $n$-tuples $\langle x_1, x_2, \ldots, x_n \rangle$, where $0 \leq x_1 \leq x_2 \leq \ldots \leq x_n < c$, and for each $i = 1, 2, \ldots, n$, $x_i$ are elements of sorted array p[] of size $c$. As a result, the function returns the number of semi-canonical elements. In her work, she will refer to function bool check(int[], int). This function refers to the use of each received $n$-tuple. It examines whether this is a semi-canonical element and prints it if the answer is yes.

```
bool check(int x[], int n, int k)
{
    int yj;     // the integer representing column (n-j)
    int y0=0;   // the integer preceding column j
    int b;
    for (int j=n-1; j>=0; j--)
    {
        yj=0;
        for (int i=0; i<n; i++)
        {
            b = 1<<j & x[i] ? 1 : 0;
            yj |= b << (n-1-i);
        }
        if (yj<y0 || (NumbOf_1(yj) != k)) return false;
        y0 = yj;
    }
// We have received a canonical element. Print it:
    for (int i=0; i<n; i++) cout<<x[i]<<" ";
    cout<<'\n';
    return true;
}

int n_tuple(int p[], int n, int k, int c)
{
    int t=0;
    int a[n], x[n];
```

```
    int indx = n-1;
    for (int i=0; i<n; i++) a[i]=0;
    while (indx >= 0)
        {
        for (int i=indx+1; i<n; i++)   a[i] = a[indx];
        for (int i=0; i<n; i++)   x[i] = p[a[i]];
        if(check(x,n,k)) t++;
        indx = n-1;
        a[indx]++;
        while (indx>=0 && a[indx]==c)
            {
            indx--;
            a[indx]++;
            }
        }
    return t;
}
```

Here we will omit the description of the function main().

## 5. Conclusion – some results from the work of the algorithm

Let $n$ and $k$ be natural numbers. Let us denote with $\nu(n,k)$ the number of all semi-canonical matrices in $\Lambda_n^k$, i.e.

$$\nu(n,k) = \left|\mathcal{Z}_n^k\right| = \left|\Lambda_n^k \cap \mathcal{V}_n\right|.$$

Here for $k = 2, 3, 4$ and $5$ we will indicate the beginning of the sequences $\{\nu(n,k)\}_{n=k}^{\infty}$ for some not big values of the parameter $n$. With the help of the computer program described in section 4, the following results have been obtained:

$$\{\nu(n,2)\}_{n=2}^{\infty} = \{1,1,2,5,13,42,155,636,2889,14321,76834,443157,\ldots\} \quad (5)$$

$$\{\nu(n,3)\}_{n=3}^{\infty} = \{1,1,3,25,272,4070,79221,1906501,\ldots\} \quad (6)$$

$$\{\nu(n,4)\}_{n=4}^{\infty} = \{1,1,5,161,7776,626649,\ldots\} \quad (7)$$

$$\{\nu(n,5)\}_{n=5}^{\infty} = \{1,1,8,1112,287311,\ldots\} \quad (8)$$

The integer sequences (5), (6), (7) and (8) are indicated in the Encyclopedia of Integer Sequences [1], respectively under the numbers A229161, A229162, A229163 and A229164. All of them are presented by N. J. A. Sloane, who cites the work [9].

The sequence (5) is commented by Brendan McKay and is supplemented by R. H. Hardin with the elements $\nu(12,2) = 76\,834$ and $\nu(13,2) = 443\,157$.

In the sequence (6) the element $\nu(10,3) = 1\,906\,501$ is added by R. H. Hardin.

## References

[1] The On-Line Encyclopedia of Integer Sequences (OEIS). http://oeis.org/
[2] I. Bouyukliev, "About Algorithms for Isomorphism-Free Generations of Combinatorial Objects," Mathematics and education in mathematics, (38), pp. 51-60, 2009.
[3] [3] H. Kostadinova, K. Yordzhev "A Representation of Binary Matrices," Mathematics and education in mathematics, (39), pp. 198-206, 2010.
[4] [4] V. N. Sachkov, V. E. Tarakanov, Combinatorics of Nonnegative Matrices of Nonnegative Matrices, Amer. Math. Soc., 1975.
[5] [5] V. E. Tarakanov, Combinatorial problems and (0,1)-matrices. Moscow, Nauka, 1985 (in Russian).
[6] [6] K. Yordzhev, "An example for the use of bitwise operations in programming," Mathematics and education in mathematics, (38), pp. 196-202, 2009.
[7] [7] K. Yordzhev, "Bitwise Operations Related to a Combinatorial Problem on Binary Matrices," I. J. Modern Education and Computer Science, (4), pp. 19-24, 2013.
[8] [8] K. Yordzhev, The Bitwise Operations Related to a Fast Sorting Algorithm. International Journal of Advanced Computer Science and Applications (IJACSA), Vol. 4, No. 9, 2013, pp. 103-107.
[9] [9] K. Yordzhev, "Fibonacci sequence related to a combinatorial problem on binary matrices," preprint, arXiv:1305.6790, 2013.

## AUTHOR

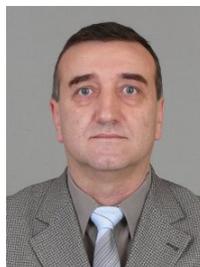

**Associate professor Dr. Krasimir Yordzhev** is a lecturer in computer science, programming and discrete mathematics at the Department of Computer Science, Faculty of Mathematics and Natural Sciences, South-West University, Blagoevgrad, Bulgaria. Dr. Yordzhev received his PhD degree in the Faculty of Cybernetics, Kiev State University, Ukraine. He is the author of more than 70 scientific publications in the field of discrete mathematics, combinatorics, combinatorial algorithms and programming.